\documentclass[preprint,12pt]{elsarticle}

\usepackage{psfrag}
\usepackage[T1]{fontenc}
\usepackage{color}
\usepackage{times}
\usepackage{amsmath}
\usepackage{amssymb}
\usepackage{graphicx}
\usepackage{wasysym}
\usepackage{textcomp}

\begin{document}

\begin{frontmatter}

  \title{Holographic interpretation of two-dimensional O($N$) models
    coupled to quantum gravity}

\author{Wolfhard Janke}
\author{Adriaan M. J.~Schakel}
\address{Institut f\"ur Theoretische Physik, Universit\"at Leipzig,
Postfach 100 920,
D-04009 Leipzig, Germany}
\begin{abstract}
  Various two-dimensional O($N$) models coupled to Euclidean quantum
  gravity, whose intrinsic dimension is four, are shown to belong to
  universality classes of nongravitating statistical models in a lower
  number of dimensions.  It is speculated that the matching critical
  behaviors in the gravitating and dimensionally reduced models may be
  manifestations of the holographic principle.
\end{abstract}


\end{frontmatter}

\section{Introduction}
\begin{figure}
\centering
\includegraphics[width=0.3\textwidth]{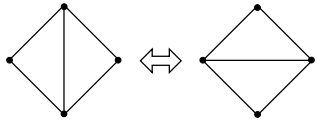}
\caption{Pachner move that flips a bond shared by two equilateral triangles.  
The move involves deleting a bond and replacing it with a new one.
\label{fig:triangulation}}
\end{figure}

The critical behavior of various two-dimensional (2D) statistical models
coupled to quantum gravity with an Euclidean signature is known exactly.
These exact results are provided in part by the
Knizhnik-Polyakov-Zamolodchikov (KPZ) formula \cite{KPZ}, which
transcribes the critical exponents of 2D statistical models into the
exponents characterizing the critical behavior of these models when
coupled to gravity.  For our purpose it is convenient to regularize
these theories of gravity by putting the 2D statistical models on a
fluctuating surface constructed through, for example, dynamical
triangulation.  In its simplest form, the surface is built from
equilateral triangles \cite{Kazakov_tri,Ambjorn,David_tri}.  A canonical
ensemble of random surfaces obtains by gluing a given number of
(identical) triangles in all possible ways to form a compact surface of
fixed (say spherical) topology.  Such an ensemble can be sampled in a
Monte Carlo simulation by using the (local) Pachner move \cite{Pachner}
shown in Fig.~\ref{fig:triangulation} as update.  This update, which
consists of a simple bond flip, is both ergodic and preserves the
topology of the surface.  On the lattice, the functional integral over
geometries in the continuum theory is replaced with a sum over all
triangulations, each carrying the same weight.  Fluctuations in the
geometry are in this way properly accounted for in the lattice model.
It turns out that the fluctuations are so large that the surface becomes
highly irregular and cannot usually be embedded in 3D space.  In fact,
the fluctuating surface, representing pure quantum gravity, has the
large fractal, or Hausdorff dimension $d=4$ \cite{KKMW}.

Matter can be included on such an empty fluctuating surface by
introducing fields (spin variables) on, for example, the lattice sites.
Usually only nearest neighbor interactions are considered for
convenience, i.e., only lattice sites connected by a bond are assumed to
interact.  The matter fields and the lattice must be updated
simultaneously, taking into account that a Pachner move changes the
nearest neighbor contingency table.  The fractal dimension of the
decorated fluctuating surface is not known exactly, but numerically
found to be consistent with $d=4$ for matter fields with central charge
$0 \leq c \leq 1$ \cite{cneq0}.  Given this observation, it seems
somewhat deceptive to refer to these models as ``two-dimensional''
quantum gravity, as is commonly done.  This seems even more so in light
of the fact that it is the fractal dimension of the fluctuating surface
that features in the scaling laws (see below) and takes the place of the
dimensionality of a regular lattice in the absence of gravity.  Below,
we take the point of view that, at least for $0 \leq c \leq 1$, the
gravitating statistical models are in $d = 4$. In line with the usual
application of the O($N$) model to thermal phase transitions in
equilibrium, for which time is irrelevant and can be ignored, we take
the dimensions to represent space.

The critical exponents of the 2D Ising model on a fluctuating planar
surface were first determined exactly by Boulatov and Kazakov
\cite{Kazakov}, who came to the surprising conclusion that these
exponents are identical to those of the spherical model on a regular 3D
lattice.  This match in critical behavior is commonly considered a mere
coincidence.  By providing additional matches of this type, we in this
paper speculate that, instead of being coincidences, they may be
manifestations of the \emph{holographic} principle \cite{tHooft}.

The paper is organized as follows.  We start by recalling some key
results of the O($N$) spin model on a flat planar surface which will be
used in Sec.~\ref{sec:fl} to study the model on a fluctuating surface.
We show that, in addition to the Ising model ($N=1$), for two other
values of $N$, the critical exponents of the O($N$) model on a
fluctuating lattice are mirrored by a statistical model defined on a
regular lattice.  In Sec.~\ref{sec:tri_O}, we extend the analysis to the
tricritical O$\left(N^\mathrm{t}\right)$ model and uncover an additional
match in critical behavior.  The nongravitating mirror models are all
in 3D or 2D, i.e., they are of lower dimension than their 4D gravitating
counterparts.

\section{O($N$) model}
\label{sec:O}
The O($N$) spin model on a flat planar lattice with intrinsic
dimension $d=2$ undergoes for $-2 \leq N \leq 2$ a continuous
phase transition \cite{Nienhuis_rev}.  The model, parametrized as
\begin{equation}
\label{N} 
N = 2 \cos \left(\frac{\pi}{m} \right)
\end{equation} 
with $1\leq m \leq \infty$, has central charge $c = 1 -
6/m(m+1)$, and thermal and magnetic exponents \cite{Nienhuis}
\begin{equation} 
\label{ce}
\frac{1}{\nu d} = 1 - \Delta_E, \quad \frac{2 -
  \eta}{d} = 1 - 2 \Delta_1 .
\end{equation} 
The remaining critical exponents, $\beta, \gamma$, and $\delta$
follow from the two independent exponents $\nu$ and $\eta$ through
scaling relations.  In Eq.~(\ref{ce}), $\Delta_E = \Delta_{1,3}$ and
$\Delta_1 = \Delta_{1/2,0}$ denote conformal dimensions expressed in
terms of entries in the Kac table \cite{Kac}
\begin{equation} 
\label{kac}
\Delta_{p,q} = \frac{[(m+1)p -m q]^2 -1}{4m(m+1)} .
\end{equation} 
The conformal dimension $\Delta$ of an operator $\phi$ determines the
scaling dimension $x$ of that operator, which in turn specifies the
algebraic decay of the correlation function at the critical point $
\langle \phi(\mathbf{r}) \phi(\mathbf{r}') \rangle \sim
1/|\mathbf{r}-\mathbf{r}'|^{2 x}$, through $\Delta = x/d$.  The
corresponding renormalization-group eigenvalue is
\begin{equation} 
  y = d (1 - \Delta) =   d - x,
\end{equation}   
with $d=2$ the dimensionality of a flat surface.

In addition to the standard spin formulation, the O($N$) model also
allows for a geometric formulation in terms of high-temperature (HT)
graphs \cite{Stanley_book}.  Contributions to the partition function are
represented in this equivalent formulation by closed graphs along the
links on the underlying lattice.  For computational convenience and
without changing the universality class, often a truncated O($N$) model
is considered \cite{DMNS}.  Whereas a link in the standard HT expansion
can be multiply occupied, in the truncated model it can at most be
occupied once.  A spin operator at site $\mathbf{r}$ is pictured in the
geometrical approach by putting a bond, or leg on one of the links
emanating from that site.  As a result, contributions to the spin-spin
correlation function $G_1(\mathbf{r},\mathbf{r}') \equiv \langle
\mathbf{s}(\mathbf{r}) \cdot \mathbf{s} (\mathbf{r}') \rangle$ feature,
in addition to possible closed graphs, an open graph, or strand,
connecting the legs at $\mathbf{r}$ and $\mathbf{r}'$.  The subscript
"1" on the conformal dimension in Eq.~(\ref{ce}), determining the
algebraic behavior of $G_1(\mathbf{r},\mathbf{r}')$, indicates that
$\Delta_1$ denotes the dimension of the one-leg operator.  An
additional operator of interest is the two-spin, or two-leg operator,
whose scaling dimension is $ \Delta_2 = \Delta_{1,0}$.  The
corresponding renormalization-group eigenvalue $y_2$ determines the
fractal, or Hausdorff dimension $D_\mathrm{HT}$ of the HT graphs of the
O($N$) model:
\begin{equation} 
\label{DHT}
\frac{D_\mathrm{HT}}{d} = \frac{y_2}{d} = 1 - \Delta_2 .
\end{equation} 
A final operator of interest in the geometrical approach is the four-leg
operator, which introduces intersections in HT graphs.  Its conformal
dimension is \cite{Nienhuis_rev} $\Delta_4 = \Delta_{2,0}$.  For all
$1\leq m < \infty$, the corresponding scaling dimension is larger than
two, the dimension of space, $x_4= 2 \Delta_4>2$, and $x_4= 2$ in the
limit $m \to \infty$, which corresponds to the XY model.  The four-leg
operator is therefore irrelevant at the critical point for $1\leq m <
\infty$ and becomes marginal in the limit $m \to \infty$.  Graphs at the
O($N$) critical point are for this reason referred to as \textit{dilute}
graphs, and the critical properties of the O($N$) model can be studied
on a lattice with coordination number $z=3$, where closed graphs simply
cannot intersect, without changing the universality class.  The
advantage of such a lattice is that closed graph configurations can be
uniquely decomposed into mutually and self-avoiding loops.

The one-, two, and four-leg operators belong to the set of so-called
watermelon operators with conformal dimension \cite{SD}
\begin{equation} 
\Delta_L = \Delta_{L/2,0} ,
\end{equation} 
$L=1,2, \ldots$.  The corresponding correlation function
$G_L(\mathbf{r},\mathbf{r}')$ involves $L$ HT strands connecting the
lattice sites $\mathbf{r}$ and $\mathbf{r}'$.

The fractal and conformal dimensions characterizing the O($N$) model
also appear in the context of the Potts model.  That spin model, too,
can be equivalently formulated in purely geometrical terms, this time
involving clusters.  Two types of clusters can be distinguished: plain
and so-called Fortuin-Kasteleyn (FK) bond clusters.  Plain bond clusters
are formed by unconditionally setting bonds between nearest neighbor
sites with like spins.  FK clusters, which form the basis of the
equivalent representation of the $Q$-state Potts model as a correlated
bond percolation model \cite{FK}, are constructed from the plain
clusters by erasing bonds between like spins with a prescribed,
temperature-dependent probability.  In 2D (and in two only), both types
of clusters percolate right at the critical point.  Moreover, while the
FK clusters encode in their fractal structure critical properties of the
$Q$-state Potts model, the plain clusters encode in their fractal
structure tricritical properties \cite{tri_geo}.  The latter usually
obtains when vacant sites are included in the Potts model.  If 
\begin{equation}
\label{Q} 
\sqrt{Q} = 2 \cos [\pi/(m+1)]
\end{equation}  
parametrizes the $Q$-state Potts model with central charge $c = 1 -
6/m(m+1)$, the tricritical behavior encoded in the plain bond clusters
of that model is that of the $Q^\mathrm{t}$-state tricritical Potts
model with the same central charge, parametrized as
\begin{equation}
\label{Qt}
\sqrt{Q^\mathrm{t}} = N ,
\end{equation} 
with $N$, given by Eq.~(\ref{N}), restricted to $0 \leq N \leq 2$, i.e.,
$2 \leq m \leq \infty$.  Note that this parametrization can be obtained
from that of the $Q$-state Potts model by applying the map $m \to -m
-1$, which conserves the central charge $c$.

\section{Gravitating O($N$) model}
\label{sec:fl}
We next turn to the O($N$) model on fluctuating planar lattices.  The
KPZ formula \cite{KPZ},
\begin{equation} 
  \Delta = \left(1 - \frac{m}{1+m} \right) \tilde{\Delta} +
  \frac{m}{1+m} \tilde{\Delta}^2 ,
\end{equation} 
relates the conformal dimension $\Delta$ of an operator on a flat planar
lattice to its conformal dimension $\tilde{\Delta}$ on a fluctuating
lattice, or more precisely, $\tilde{\Delta}$ is the non-negative
solution to the above equation. For the Kac table (\ref{kac}), the KPZ formula
yields the gravitationally dressed dimensions
\begin{equation} 
\tilde{\Delta}_{p,q} =  \frac{-1 + |(1+m)p - m q|}{2 m} ,
\end{equation} 
so that \cite{Kazakov,DK}
\begin{equation} 
\label{codi_fl}
  \tilde{\Delta}_E = \frac{-1+m}{m}, \quad \tilde{\Delta}_1 =
  \frac{-1 + m}{4 m}, \quad \tilde{\Delta}_2 = \frac{1}{2} ,
\quad \tilde{\Delta}_4 = 1 + \frac{1}{2 m} ,
\end{equation} 
and
\begin{equation} 
\label{ce_fl}
\tilde{\nu} d = m, \quad \frac{2 -
  \tilde{\eta}}{d} = \frac{1+m}{2 m} 
\end{equation} 
for the O($N$) model coupled to quantum gravity.  Here and in the
following, $d$ denotes the fractal dimension of the fluctuating surface.
For matter fields with central charge $0 \leq c \leq 1$, its value was
numerically found to be consistent with the value $d=4$ of an empty
fluctuating surface \cite{cneq0}.  Remarkably, the gravitationally
dressed dimensions have a simpler dependence on $m$ than their
nongravitating counterparts.  This is in particular true for
$\tilde{\Delta}_2$, which yields as fractal dimension of the HT graphs
\cite{DK}
\begin{equation} 
\label{Dtilde}
\frac{\tilde{D}_\mathrm{HT}}{d}= 1 - \tilde{\Delta}_2
= \frac{1}{2},
\end{equation}  
independent of $m$.  Such a universal behavior, where a set of models
share the same scaling dimension, is usually reserved for models in
their upper critical dimension, such as the O($N$) model on a regular 4D
lattice for which the ratio $D_\mathrm{HT}/d$ is also $\frac{1}{2}$.
The conformal dimension $\tilde{\Delta}_4$ of the four-leg operator
shows that at criticality, intersections are still irrelevant on a
fluctuating planar lattice.  This implies that the critical properties
of the O($N$) model can be studied on a fluctuating planar lattice with
coordination number $z=3$, without changing the universality class.
Also observe that in the limit $m \to \infty$, corresponding to the
gravitating XY model, the correlation length exponent $\tilde{\nu}$
diverges.  This characteristic of the XY model on a flat planar surface
is therefore preserved when the model is put on a fluctuating surface.
Finally note that for self-avoiding walks ($m=2$) on a fluctuating
random planar lattice $\tilde{\nu} = 1/\tilde{D}_\mathrm{HT}$, while for
all other values of $m$, $\tilde{\nu} \neq 1/\tilde{D}_\mathrm{HT}$.
This is similar to that found on a flat lattice \cite{ht}.

In Ref.~\cite{WM}, the fractal structure of plain and FK bond clusters
of the $Q$-state Potts model was studied on fluctuating planar lattices
with coordination number $z=3$.  The fractal dimension of the hulls of
plain bond clusters on a fluctuating lattice was, among others,
conjectured on the basis of the KPZ formula, with the result
(\ref{Dtilde}).  The predictions for the plain and FK cluster dimensions
were confirmed numerically in that reference for the Ising model ($Q=2$)
coupled to quantum gravity through Monte Carlo simulations.  A rigorous
derivation of the \emph{geometrical} KPZ relation, using a probabilistic
approach, was provided in Ref.~\cite{DuSh}.  For a derivation based on
the heat-kernel method, see Ref.~\cite{David}.

We next show that for various $N$, the O($N$) model coupled to quantum
gravity belongs to a universality class (as defined by the standard
critical exponents) of a statistical model formulated on a regular lattice,
i.e., without gravity.  Specifically:
\begin{itemize}
\item for $m=1$, the
critical exponents (\ref{ce_fl}) are the same as on a flat planar
lattice, so that putting the Gaussian model ($N=-2$) on a fluctuating
lattice has no effect on its critical behavior, 
\item for $m=\frac{4}{3}$ $\left(N= - \sqrt{2}\right)$, the critical
  exponents (\ref{ce_fl}) are mirrored by the $Q=4$ Potts model
  on a flat planar lattice,
\item for the Ising model ($m=3$) coupled to quantum gravity, the
  critical exponents (\ref{ce_fl}) are known to coincide with those of
  the spherical model on a cubic lattice \cite{Kazakov}.
\end{itemize}
In summary, for $m=1,\frac{4}{3}$, and $3$ (and surely also for other
values), the O($N$) model coupled to quantum gravity with intrinsic
dimension $d = 4$ belongs to universality classes of nongravitating
models in dimension $d=2,2$, and $3$, respectively.  Note that these
universality classes are all in lower than four dimensions---the
dimensionality of the fluctuating surface.  That is, we witness a
``dimensional reduction in quantum gravity'' \cite{tHooft}.

Although the critical exponents coincide, the fractal dimensions of the
HT graphs featuring in the gravitating and in the dimensionally reduced
models without gravity do not.  Whereas, for example, the fractal
dimension of the gravitating Gaussian model is $\tilde{D}_\mathrm{HT}/d
= \frac{1}{2}$, that of its nongravitating counterpart on a flat planar
lattice is $D_\mathrm{HT}/d = \frac{5}{8}$.  Both models are
nevertheless classified as belonging to the same universality class on
the basis that their critical exponents coincide.

\section{Tricritical O$\left(N^\mathrm{t}\right)$ model}
\label{sec:tri_O}
We next extend our analysis to the tricritical
O$\left(N^\mathrm{t}\right)$ model.  When vacant sites are included, the
O($N$) spin model, too, displays in addition to critical also
tricritical behavior.  The tricritical point is reached by gradually
increasing the activity of the vacancies.  The continuous phase
transition then eventually becomes discontinuous with the endpoint,
marking the change in the order of the transition, being the tricritical
point.  The tricritical O$\left(N^\mathrm{t}\right)$ model with central
charge $c$ is parametrized as~\cite{GNB}
\begin{equation} 
\label{nt}
N^\mathrm{t} = \sqrt{Q} - 1/\sqrt{Q} 
\end{equation} 
with $Q$ given by Eq.~(\ref{Q}), denoting the Potts model with the same
central charge, and restricted to $0 < Q \leq 2$, i.e., $1 < m \leq
\infty$.  This parametrization constitutes the tricritical counterpart
of Eq.~(\ref{Qt}), which relates the critical O($N$) model to the
tricritical $Q^\mathrm{t}$-state Potts model.

In Ref.~ \cite{ht}, we conjectured that the fractal dimension
$D_\mathrm{HT}^\mathrm{t}$ of the HT graphs at the tricritical point on a
flat planar lattice, where they collapse, follow from those (\ref{DHT}) at the
critical point  by applying the central-charge conserving map
$m \to -m -1$, so that $\Delta_2 = \Delta_{1,0} \to \Delta_{0,1} =
\Delta^\mathrm{t}_2$.  This yields
\begin{equation} 
\label{Dt}
\frac{D_\mathrm{HT}^\mathrm{t}}{d} = 1- \Delta_2^\mathrm{t} =
\frac{1+3m}{4 m} .
\end{equation}
In the limit $m\to \infty$ ($N\to 2$), the fractal dimensions
$D_\mathrm{HT}$ and $D_\mathrm{HT}^\mathrm{t}$ approach the limiting
value $\tfrac{3}{2}$ from opposite directions, with the collapsing
graphs being, of course, more crumpled than the dilute graphs.  In that
limit, corresponding to $c=1$, the critical and tricritical fixed points
merge.  In the opposite limit, $m \to 1$, the collapsing graphs even
start to fill the entire available space, $D_\mathrm{HT}^\mathrm{t}/d
\to 1$.  Equation~(\ref{Dt}) reduces to the known results
$D_\mathrm{HT}^\mathrm{t}= \frac{7}{4}$ for polymers ($m=2, \;
N^\mathrm{t}=0$) at the theta point \cite{Coniglioetal}, and
$D_\mathrm{HT}^\mathrm{t}= \frac{13}{8}$ for the tricritical Ising model
($m=4, \; N^\mathrm{t}=1$), which coincides with the tricritical
$Q^\mathrm{t}=2$ Potts model.  We further conjectured in Ref.~\cite{ht}
that also the dimensions of the leading magnetic operators of the
critical and tricritical models are related by the dual map $m \to
-m-1$.  Specifically, $\Delta_1 = \Delta_{1/2,0} \to \Delta_{0,1/2} =
\Delta_1^\mathrm{t}$.  This prediction has been confirmed by
high-precision numerical transfer-matrix calculations
in\footnote{Earlier numerical work by Guo \emph{et
    al.}~\cite{bloeteetal} already showed agreement with our prediction
  for $0 \leq c \lesssim 0.7$, but the two data points reported in the
  interval $0.7 \lesssim c \lesssim 1.0$ deviated significantly from our
  prediction (see Fig.~1 of Ref.~\cite{ht}).  In Ref.~\cite{GNB}, the
  data point $\{c,x_h\}=\{0.860(1), 0.0923(2)\}$ of the earlier work
  shifted upwards to $\{0.860(1),0.0950(2)\}$, while the old data point
  $\{0.998(2),0.111(1)\}$ received a much larger error bar
  $\{1.001(2),0.12(1)\}$, so that the revised numerical results now fit
  our prediction perfectly over the entire range $0 \leq c \leq 1$.
  Moreover, it was shown that the transition of the tricritical model
  ceases to be continuous for $c>1$, in accordance with our prediction.}
Ref.~\cite{GNB}.

Transcribing the conjecture (\ref{Dt}) to the tricritical
O$\left(N^\mathrm{t}\right)$ model on a fluctuating planar lattice, we
obtain the fractal dimension
\begin{equation} 
  \frac{\tilde{D}_\mathrm{HT}^\mathrm{t}}{d} = 1 - 
  \tilde{\Delta}_2^\mathrm{t} =  \frac{1+m}{2m} ,
\end{equation}    
which, in contrast to the fractal dimension of dilute graphs on a
fluctuating lattice, does depend on $m$.  As expected, the collapsing
graphs are more crumpled than the dilute graphs for all $1 < m <
\infty$, see Eq.~(\ref{DHT}).  In the limit $m \to 1$, the collapsing
graphs fill the entire available space,
$\tilde{D}_\mathrm{HT}^\mathrm{t}/d \to 1$, as on a flat planar lattice.
The two fractal dimensions $\tilde{D}_\mathrm{HT}/d$ and
$\tilde{D}_\mathrm{HT}^\mathrm{t}/d$ coincide in the limit $m \to
\infty$.  The KPZ formula yields as magnetic dimension for the
gravitating tricritical O$\left(N^\mathrm{t}\right)$ model
\begin{equation} 
\label{xtilde}
\tilde{\Delta}_1^{\,\mathrm{t}} = \frac{-2 + m}{4m} ,
\end{equation} 
which, as on a flat planar lattice, is physical only for $m \ge 2$.
With the exception of collapsing self-avoiding walks and of the
tricritical Ising model, the leading tricritical thermal exponent on a
flat planar lattice does not correspond to an entry in the Kac table
\cite{GNB}.  When conformal dimensions are not rational numbers, it is
not clear whether these can be fed into the KPZ formula.  In the
following, we, therefore, restrict ourselves to the gravitating
O$\left(N^\mathrm{t}=0\right)$ and O$\left(N^\mathrm{t}=1\right)$
models.

For collapsing self-avoiding walks \cite{DStheta} ($m=2, \;
N^\mathrm{t}=0$) the conformal dimension of the leading thermal operator
is $\Delta^\mathrm{t}_E = \Delta_{2,2} = \frac{1}{8}$, while for the
tricritical Ising model \cite{Zamolodchikov} ($m=4, \; N^\mathrm{t}=1$)
it is $\Delta^\mathrm{t}_E = \Delta_{1,2}=\frac{1}{10}$.  Both 
dimensions happen to yield $\tilde{\nu}^\mathrm{t} d = \frac{4}{3}$
for the tricritical correlation length exponent on a fluctuating planar
lattice.  Together with the magnetic dimension (\ref{xtilde}), this
gives for the exponents of collapsing self-avoiding walks coupled to
quantum gravity:
\begin{equation} 
  \frac{1}{\tilde{\nu}^\mathrm{t} d} = 1 -
    \tilde{\Delta}^\mathrm{t}_E = \frac{3}{4}, \quad \frac{2 -
    \tilde{\eta}^\mathrm{t}}{d} = 1 - 2
  \tilde{\Delta}^\mathrm{t}_1 = 1 ,
\end{equation} 
and for the gravitating tricritical Ising model:
\begin{equation} 
\frac{1}{\tilde{\nu}^\mathrm{t} d} = \frac{3}{4}, \quad
\frac{2 -
    \tilde{\eta}^\mathrm{t}}{d} = \frac{3}{4} .
\end{equation}  
The remaining tricritical exponents $\tilde{\beta}^\mathrm{t},
\tilde{\gamma}^\mathrm{t}$, and $\tilde{\delta}^\mathrm{t}$
follow from $\tilde{\nu}^\mathrm{t}$ and
$\tilde{\eta}^\mathrm{t}$ through the usual scaling relations.  Note
that for collapsing self-avoiding walks coupled to gravity,
$\tilde{\nu}^\mathrm{t} = 1/\tilde{D}_\mathrm{HT}^\mathrm{t}$,
so that the characteristic relation between correlation length exponent
and fractal dimension for (collapsing) self-avoiding walks on flat
planar lattices persists on fluctuating lattices.

The gravitating tricritical Ising model also belongs to a universality
class defined by a nongravitating model in lower dimensions.
Specifically, it shares the same critical exponents as the fourth-order
critical point of the Ising model on a cubic lattice with
antiferromagnetic nearest neighbor and ferromagnetic next-nearest
neighbor interactions in an external magnetic field \cite{KiCo}.
Although the exponents determined in Ref.~\cite{KiCo} are mean-field
exponents, we expect them to be exact in $d = 3$, for this is above the
upper critical dimension
\begin{equation} 
d_\mathrm{u} = \frac{2k}{k-1} = \frac{8}{3}
\end{equation} 
of a fourth-order critical point ($k=4$).  That is, we again witness
``dimensional reduction in quantum gravity'' \cite{tHooft}, with the
gravitating tricritical Ising model in $d = 4$ sharing the same critical
exponents as a nongravitating model in $d = 3$.

\section{Discussion}
\label{sec:discussion}
In this paper, the critical properties of the 2D O($N)$ model coupled to
Euclidean quantum gravity, whose intrinsic dimension is four, were
studied.  It was pointed out that for various $N$, the critical
exponents of these models are mirrored by nongravitating models in
dimensions lower than four.  Rather than consider this to be mere
coincidence, we submit that these matches may, in fact, be
manifestations of the holographic principle.  This principle, put
forward by 't~Hooft \cite{tHooft}, asserts that all of the information
about a system with gravity in some region of space, which can be very
large, or even infinite, is coded in degrees of freedom that are
confined to a boundary to that region and are not subject to gravity.
At the least, the principle mandates that the critical properties of a
statistical model coupled to gravity be mirrored by one without gravity
in one dimension less, representing the gravity-free degrees of freedom
on a boundary to space.  Since such matches in critical behavior exist
for the gravitating critical and tricritical Ising models, it is
tempting to speculate that these models are holographic projections of
the corresponding nongravitating models defined on a 3D boundary to 4D
space.

For the gravitating O$\left(N= -2\right)$ and O$\left(N= -
  \sqrt{2}\right)$ models, whose matter fields have a negative central
charge, the nongravitating counterparts are in 2D.  If the
dimensionality of the fluctuating surface with matter fields of negative
central charge is still four, and if the holographic principle applies,
it would imply that, in these cases, a 2D model suffices to reconstruct
the critical behavior of a 4D gravitating world.

\section*{Acknowledgement} 
This work is supported in part by the Deutsche Forschungsgemeinschaft
(DFG) under grant No.~JA483/23-2 and the EU RTN-Network `ENRAGE':
``Random Geometry and Random Matrices: From Quantum Gravity to
Econophysics'' under grant No.~MRTN-CT-2004-005616.

\end{document}